\documentclass[12pt,a4paper,fleqn]{article}
\usepackage{tikz}
\usepackage{caption}
\usepackage{lscape}

\usepackage{fullpage}
\usepackage{amsmath}
\usepackage{amsthm}
\usepackage{amssymb}
\usepackage{amscd}
\usepackage{plain}
\usepackage{graphicx}
\usepackage{t1enc,epsfig}
\usepackage[mathscr]{eucal}
\usepackage{epstopdf}
\usepackage[german,english]{babel}
\usepackage{stmaryrd}

\usepackage{xcolor}

\usepackage{wrapfig}
\usepackage{float}

\numberwithin{equation}{section}

\newenvironment{proof1}{{\bf Proof. }}{\hfill$\rule{1ex}{1ex}$\par\medskip}

\newtheorem*{thm}{Theorem}

\newtheorem*{prop}{Proposition}

\newtheorem*{lem-1}{Lemma 1}
\newtheorem*{lem-2}{Lemma 2}

\newtheorem*{ra}{Remark 1}
\newtheorem*{rb}{Remark 2}
\newtheorem*{rc}{Remark 3}

\begin{document}

\def\bbGam{\bbGamma}
\def\del{\delta} 
\def\s{\phi}
\def\f{\varphi}
\def\t{\tau} 
\def\e{\varepsilon}
\def\r{\varrho}
\def\Om{\Omega} 
\def\p {\partial}

\def\D{{\bf D}}

\def\tO{B} 
\def\tg{B} 
\def\bs{\overline \phi}
\def\Ev{{\cal E}}

\def\bbZ{{\mathbb Z}} 
\def\bbA{{\mathbb A}} 
\def\bbB{{\mathbb B}} 
\def\bbG{{\mathbb G}} 
\def\bbD{{\mathbb D}} 
\def\bbF{{\mathbb F}} 
\def\bbH{{\mathbb H}} 
\def\bbL{{\mathbb L}} 
\def\bbN{{\mathbb N}} 
\def\bbO{{\mathbb O}} 
\def\bbR{{\mathbb R}} 

\def\bo {{\mathbf o}}

\def\bt {{\mathbf t}}
\def\bu {{\mathbf u}}
\def\bv{{\mathbf v}}
\def\bw{{\mathbf w}}
\def\bx{{\mathbf x}}
\def\by{{\mathbf y}}
\def\bz {{\mathbf z}}

\def\re{{\rm e}} \def\ro{{\rm o}} \def\rST{{\rm ST}} 

\def\diam {{\rm diam}}

\def\bull{\bullet}
\def\cc{\circ}

\def\E{\epsilon}
\def\I{\iota}
\def\Int {I}
\def\Ext {E}

\def\co{{\theta}}

\def\wh{\widehat} \def\wt{\widetilde} \def\ov{\overline} \def\un{\underline} 
\def\cl{\centerline}

\title{\bf A Nearest-Neighbor Hard-Core Model on a Penrose Graph}

\author{\bf A. Mazel$^1$, I. Stuhl$^2$, Y. Suhov$^{2,3}$}

\date{}
\footnotetext{2010 {\em Mathematics Subject Classification:\; primary 60G60, 82B20, 82B26}}
\footnotetext{{\em Key words and phrases:} Nearest-neighbor hard-core model, particle activity, extreme Gibbs measure, ground state, contour, polymer expansion, uniqueness, Penrose P3 tiling, substitution, self-similarity, $k$-level supertile, mutually locally derivable, $k$-atlas.

\noindent
$^1$ AMC Health, New York, NY, USA;\;\;
$^2$ Math Dept, Penn State University, PA, USA;\;\;
$^3$ DPMMS, University of Cambridge and St John's College, Cambridge, UK.}

\maketitle

\begin{abstract} We prove that the maximal graph-density of an independent set in a Penrose P3 tiling 
considered as a planar non-directed graph is equal to $(57 - 25 \sqrt{5})/2 \approx 0.54915$ despite the fact that the graph is bipartite. Accordingly, the extreme Gibbs measure of the nearest-neighbor hard core particle model on this graph is unique for sufficiently large values of the particle activity. This invalidates a natural expectation to observe the coexistence of even and odd phases. 
\end{abstract}

\section{Introduction}\label{Section1} 

This paper has been triggered and inspired by the work \cite{[CHP]} where the authors consider the high activity phase diagram of the nearest neighbor hard-core models on bipartite graphs. A  connected infinite bipartite graph $\bbG$ having a countable set of vertices is naturally suitable for nearest neighbor hard-core models of statistical mechanics. The set of vertices in $\bbG$ and the set of edges in $\bbG$ are denoted by $V(\bbG)$ and $E(\bbG)$ respectively. We frequently refer to $\bbG$ rather than to $V(\bbG)$ and $E(\bbG)$ if it does not create an ambiguity.  The set $V(\bbG)$ is partitioned into even and odd parts $\bbG^\re $, $\bbG^\ro $ such that any edge of the graph connects two vertices of a different parity. A configuration $\phi$ is an element of $\{0,1\}^{V(\bbG)}$. Accordingly, the vertex $\bv\in \bbG$ with $\phi(\bv)=1$ is called {\it occupied} or containing a {\it particle} and it is called {\it vacant} or {\it empty} otherwise. A configuration $\phi$ is called {\it admissible} if it does not contain edges $(\bu,\bv)$ with $\phi(\bu)=\phi(\bv)=1$,
which is a hard-core condition at the graph distance 1. The model is defined  
on the space of admissible configurations via the following formal Hamiltonian 
%$$H(\phi)=-\log( u_e)\sum_{\bv \in \bbG^\re } \phi(\bv)-log( u_o)\sum_{\bv \in \bbG^\ro } \phi(\bv), \eqno{(1)}$$
$$H(\phi):=-\log( u)\sum_{\bv \in \bbG} \phi(\bv), \eqno{(1)}$$ where $u >0$ stands for activity. Consequently, for $u>1$ Hamiltonian (1) strongly favors as many occupied vertices as possible. Such models are classical subjects of statistical mechanics: they are important in both the fundamental and applied perspectives. We refer the reader to \cite{[CHP]} for a general exposition of the related background from statistical mechanics.

Paper \cite{[CHP]} introduces natural requirements to the graph regularity and uniformity which ensure the coexistence of even and odd phases (extreme Gibbs measures) for sufficiently large 
$u$. It is not clear to what extent the uniformity requirement is necessary. %One would think that 
%{\color{red}it} can be relaxed and replaced, say by the graph recurrence (e.g., see [KPZ]). 
The current paper demonstrates that even a uniform recurrence is sometimes not enough to guarantee the coexistence of even and odd phases.

Specifically, we consider $\bbG$ which is the set of vertices and edges of a Penrose P3 tiling of the entire $\mathbb{R}^2$ \cite{[P]}. Such a tiling is constructed from unit thin and thick rhombi (see Figure 1). The corresponding graphs $\mathbb{G}$ are bipartite \cite{[FSP]} and linearly repetitive %Sect. 10.5.4 in \cite{[GS]},  %pp 563-564, 
\cite{[FL]}.

\begin{center}
\begin{figure}[H]\label{Penrose}
\includegraphics[scale=0.62]{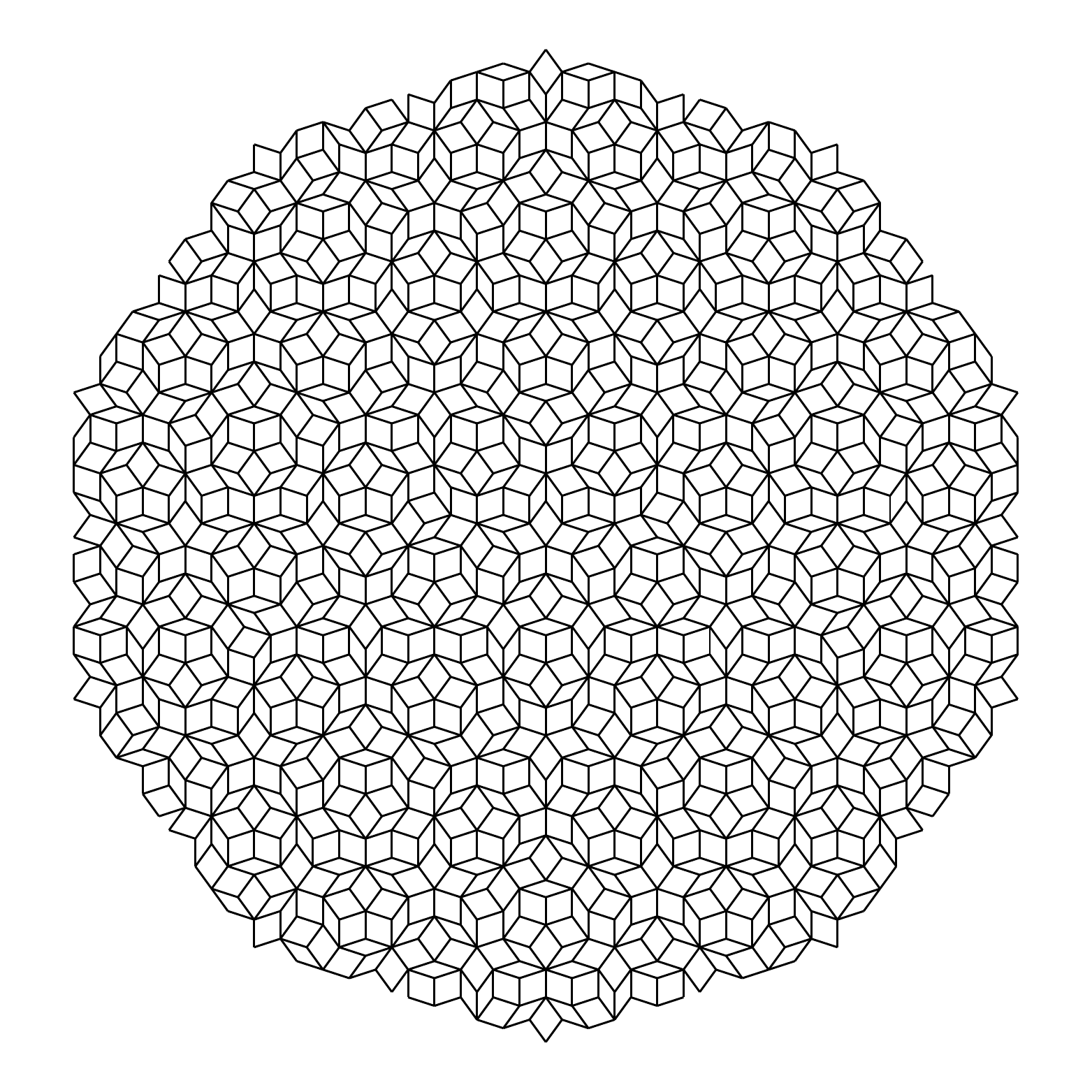}

\caption{A fragment of a  P3 tiling}  %Penrose
\end{figure} 
\end{center}

The corresponding proportion of even and odd vertices in a P3 tiling
equals $1/2$, so when $u$ is 
large enough, one would expect a familiar picture of coexistence of even and odd phases.
 A surprising result of the current paper is that the Gibbs measure turns out 
to be unique for all $u>\ov{u}$, where constant $\ov{u}\gg 1$. The particle density 
in this measure is $>1/2$. On the other hand,
it is clear that the Gibbs measure is also unique for $u <\un{u}$, where constant 
$0<\un{u}<1$. It would be interesting to analyze the phase diagram of the model for $\un{u}<u<\ov{u}$ where the employed polymer expansion technique is not applicable.

A peculiarity of a Penrose P3 graph is that it contains vertices of degrees $3$ to $7$, with the average degree $4$. Therefore, some areas of the graph have excessive vertex density 
in $\bbR^2$, intermittently for even and odd vertices. It hints at the possibility that the ground state could be a mixture of occupied even and odd vertices 
rather than uniformly occupied $\bbG^\re $ or $\bbG^\ro $. It turns out that the ground state is essentially a collage of finite patches that are fully occupied intermittently by even and odd particles. Such patches are separated by finite loops consisting of vacant edges (both endpoints are vacant). As a consequence of such intermittency, the corresponding density of occupied vertices is distinctively larger than $1/2$. Cf. Figure 2.

The densest `local' configuration of vertices of the same parity in a P3 tiling is given by three 
vertices placed along the short diagonals of two adjacent thin rhombi. The configuration where
all such triples of vertices are occupied is admissible but not saturated. Its densest admissible 
extension is a uniquely defined saturated configuration shown in Figure 2. It turns out 
that this configuration is a unique ground state; its formal construction is given in Section \ref{Section3}.

In Figure~2 the vacant edges are painted yellow. They form boundaries (loops) separating patches occupied by particles of different parity.   
Observe that Figure~2 demonstrates two distinct finite simply $2$-connected {\it patterns} of occupied vertices of the same parity completely surrounded by yellow loops. The remaining regions of the same parity can be arbitrarily large, but one can cut them across the narrow bridges where the opposite yellow boundaries are closest to each other.  Up to Euclidean shifts and rotations, the cutting produces three additional patterns. The crucial property behind the model behavior is that, for each of the five patterns, the corresponding densest configuration of particles inside the pattern does not depend on the boundary condition outside the pattern. This immediately implies that the unique
ground state is given by the concatenation of the densest configurations inside 
each pattern.

In particular, it is easy to construct a configuration with particle density exceeding~
$1/2$. Indeed, start with the fully occupied $\bbG^\re $, locate an odd patch
 congruent to the smallest of the two described patterns (see the ``urchin'' in Figure~3), and change the parity 
of the occupied vertices inside this patch. The procedure increases the total number of particles 
by 1. The density of a given pattern in a P3 tiling is equal to $(47- 29\varphi )/5$, 
where 
$$\varphi =\dfrac{1 +\sqrt{5}}{2}$$ 
is the golden ratio (cf. Figure~8 in \cite{[FL]}). Flipping the parity of particles in each such odd patch, we obtain the configuration with the overall particle density
exceeding $0.5077$. The particle density of the ground state shown in Figure~2 is evidently 
larger (see Theorem below); it turns out that it equals $(57 - 25 \sqrt{5})/2 \approx 0.54915$. 
Some related results were obtained in \cite{[FSP]} regarding the densest
dimer configuration on a P3 tiling. Our Figure~2 is an analogue of FIG.~7 in \cite{[FSP]}.

\begin{center}
\begin{figure}[H]
\hskip-30pt\includegraphics[scale=0.7]{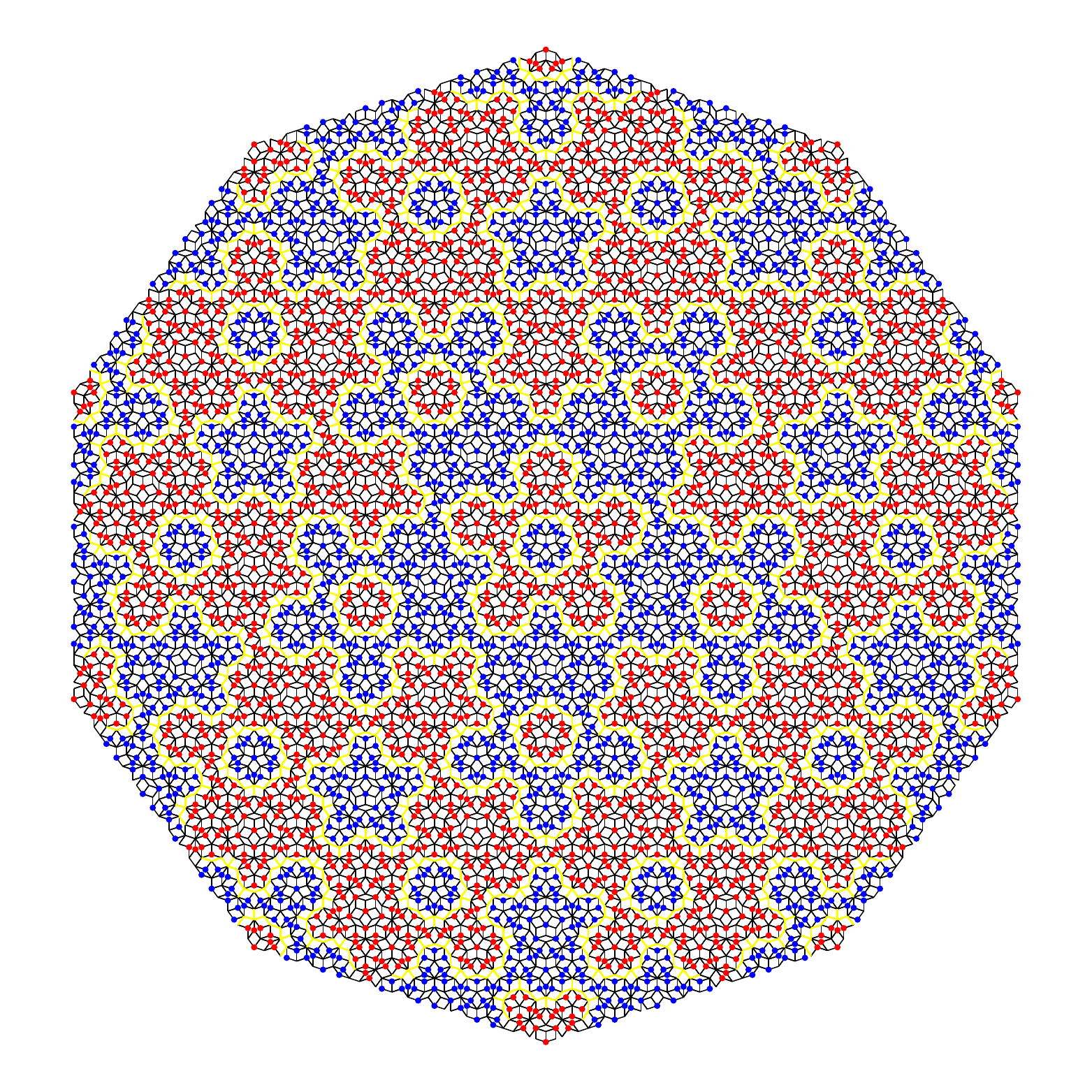}

\caption{A fragment of the ground state for a P3 tiling}
\medskip
The even occupied vertices are marked as blue dots, the odd occupied vertices are marked as red dots, and the boundaries between blue and red 2-connected components of occupied vertices are formed by yellow edges.
\end{figure}
\end{center}

In what follows we prove that the properly defined centers of patterns form a quasiperiodic 
tiling which is {\it mutually locally derivable} (MLD) with the original P3 tiling 
\cite{[So]}. The corresponding
graph is denoted by $\bbG^\rST $ (the {\it supertiling} graph). In  the considered Gibbs distribution each pattern behaves almost independently of others (similarly to a high-temperature 
regime behavior), which explains uniqueness. Accordingly, we perform a standard coarse-graining procedure and consider a configuration in each pattern as a new spin variable. All $u$-dependent factors contribute solely 
to a single-vertex potential. Additionally, we have a pair hard-core potential along edges of 
$\bbG^\rST $ which forbids some pairs of new spins. This is a typical $m$-potential \cite{[HS]} with the
constructed ground state being the corresponding unique perfect configuration. Any 
non-ground state spin (a pattern configuration) contains at least one less occupied vertex than 
the corresponding ground state spin. The rest of the argument is standard (see the next section), since the vertex degree in $\bbG^\rST $ is at most $5$. This analysis leads to the following result.

\begin{thm}\label{Theorem}  
Let $\bbG$ be a bipartite graph corresponding to an infinite P3 tiling constructed from two types of decorated rhombi via standard matching rules. Then, for 
$u >1$, Hamiltonian~$(1)$ has a unique ground state. The ground state particle density is equal 
to $\dfrac{1}{2}(57 - 25 \sqrt{5}) \approx 0.54915$. For $u > 100000$, the model has a unique 
extreme Gibbs measure which is characterized by an absolutely convergent and exponentially decaying polymer expansion around the ground state. 
\end{thm}

\begin{ra} {\rm We stress that the lower bound $u > 100000$ is chosen to simplify the 
proofs and is not optimal.}
\end{ra}

\begin{rb} {\rm The theorem addresses {\it any} P3 tiling, regardless of its symmetry properties.
The universal construction of a P3 tiling can be found in \cite{[P]}.}
\end{rb}

\section{A coarse-grained model}\label{Section2}

The theorem is proven by a reduction of the original model to a simplified one. The target class of simplified models is described below.

Consider an infinite graph $\bbG$ with (a) countably many vertices $\bv$, (b) uniformly bounded vertex degree $d(\bv) < d <\infty$, 
(c) an isoperimetric property that for any finite connected subset of vertices $C \subset \bbG$ the number of vertices in its (internal) boundary $\partial C$ does not exceed $\max(d|C|,|C|^b)$, where $0<b<1$. Here the boundary $\partial C$ consists of vertices in $C$ having an adjacent vertex outside of $C$, and $|\cdot |$ denotes the number of vertices.

Fix a finite spin space $S$ and consider configurations $\sigma \in S^{V(\bbG)}$. For every $\bv \in V(\bbG)$ fix a subspace $S_{\bv} \subset S$  of spins allowed at $\bv$. Similarly, for every pair of adjacent vertices $(\bu,\bv) \in \bbG$ fix  a subspace $S_{\bu\bv} \subset S_{\bu} \times S_{\bv}$ 
of spin pairs allowed at  $(\bu,\bv)$. A configuration $\sigma$ is called {\it admissible} if $
\sigma(\bv) \in S_{\bv}$ and $\big(\sigma(\bu),\sigma(\bv)\big) \in S_{\bu\bv}$ for any vertex $\bv 
\in \bbG$ and any edge $(\bu,\bv)\in \bbG$.

Assume that for every $\bv \in \bbG$ and any boundary condition specified on the adjacent
vertices there exist at least $k$ allowed values of $\sigma(\bv)$, where $1 < k \le |S|$. 
(Cf. Richness Assumption in \cite{[MSS]}.) Here $|S|$
stands for the cardinality of $S$, and the assumption ensures that the space of admissible
configurations is rich enough. Consider a potential $U: S \to \mathbb{R}$ with the
following properties. For each $\bv$ the minimum of $U\big(\sigma(\bv)\big)$ over $\sigma(\bv)\in S_{\bv}$ is achieved 
at a single element $s_{\bv} \in S_{\bv}$. The configuration $\sigma(\bv)=s_{\bv},\, \bv \in 
\bbG$ is admissible. Such a configuration is called {\it perfect}. Finally, assume that
$$\min_{\sigma(\bv)\in S_{\bv}\setminus s_{\bv}} U\big(\sigma(\bv)\big) - U(s_{\bv}) > \tau, \eqno{(2)}$$
where $\tau > 0$ is independent of $\bv$. The simplified model is defined on the space of admissible configurations $\sigma$ via the formal Hamiltonian
$$H(\sigma):= \sum_{\bv \in \bbG} U\big(\sigma(\bv)\big).\eqno{(3)}$$ We use a specific procedure of  coarse graining to reduce the original model (1) to a model of type (3). The latter has an advantage that the corresponding limit Gibbs measure can be directly written in terms of convergent polymer expansions and consequently is unique.

Without loss of generality, we assume that $U(s_{\bv}) =0 $ for all $\bv \in \bbG$, i.e., the energy of the perfect configuration is $0$. A vertex $\bv$ is called {\it correct} in an admissible configuration $\sigma$ if $\sigma(\bv)=s_{\bv}$ and $\sigma(\bu)=s_{\bu}$ for all adjacent 
vertices $\bu$. A connected component of the set of the vertices which are not correct is referred to as a {\it contour} 
of configuration $\sigma$. Formally, a contour $\co$   is a pair $\big(C,\sigma(C)\big)$ consisting 
of a connected set of vertices $C$ called a contour support (or a support, for short) and a
configuration $\sigma(C)$ in $C$. The statistical weight of a contour $\co$ with a finite support 
$C$ is equal to
$$\begin{array}{cl}w(\co) &:= \exp\Big(-H\big(\sigma(C)\big)\Big)\\
\;&:= \exp \left( -\sum\limits_{\bv \in C} U\big(\sigma(\bv)
\big)\right) = \exp \left( -\sum\limits_{\bv \in C:\, \sigma(\bv)\not= s_{\bv}} 
U\big(\sigma(\bv)\big)\right).\end{array}\eqno{(4)}$$
As a result, we construct a one-to-one map between compatible collections of contours 
$\{\co_i\}$ and admissible configurations $\sigma$ such that the statistical weight of a
configuration is a product of the statistical weights of its contours. The underlying compatibility 
condition for contours is that their supports are not connected.

By construction, for at least $|C|/(d+1)$ vertices $\bv \in C$ inside the support of a contour 
$\co = \big(C,\sigma(C)\big)$ one has $\sigma(\bv) \not=s_{\bv}$, and consequently
$$w(\co) \le \exp\left(-{\tau \over d+1}|C|\right).\eqno{(5)}$$
On the other hand, $|S|^n\, (ed)^n$ is a standard upper bound for the number of contours 
$\co = \big(C,\sigma(C)\big)$ whose supports $C$ contain a given vertex $\bo$ and have 
size $|C|=n$.  

Now we are ready to employ the polymer expansions. The subject of polymer expansions is 
rather wide and contains a large number of original papers.  Instead of them we are using 
the appendix in \cite{[KLMS]} as the reference most tailored to our purposes. Note that in our model a 
polymer is simply a finite collection of contours whose union is a connected subset of $\bbG$.

\begin{prop}
Adopt the above assumptions on the coarse-grained model under consideration. Suppose that $\tau > 10(d+1) \log(|S|ed)$ and consider a finite 
sub-graph (volume) $\Lambda \subset \bbG$ with an arbitrary admissible configuration in 
$\bbG\setminus \Lambda$ serving as a boundary condition.  Then the logarithm of the corresponding  
partition function for Hamiltonian {\rm (3)} can be represented as an absolutely convergent sum of the 
statistical weights of the polymers of contours belonging to $\Lambda$. The emerging limit Gibbs 
measure (with an arbitrary sequence of boundary conditions) is characterized by the probabilities 
of the cylinder events explicitly represented in terms of statistical weights of contours and 
polymers. The obtained Gibbs measure manifests an exponential decay of truncated 
correlation functions, and is unique for the above range of $\tau$.
\end{prop}

\begin{proof1} In the course of this proof references (A.n) point to the displayed equations in 
 \cite{[KLMS]}. Accordingly, the one-to-one map between configurations and compatible collections of 
contours implies the representation (A.1) for the partition function in a finite  volume $\Lambda$ 
with an arbitrary boundary condition.  This representation remains intact if one unites all contours 
with given support into a factor contour. The statistical weight of a factor contour is simply the 
sum of statistical weights of the corresponding original contours. From now on we deal with factor 
contours only and reassign to them the term contour.

Condition (A.3) of Theorem~A.1 holds true for $a(\theta)= |C|/10$ and $\tau > \tau_0 
= (d+1)\log(13ed|S|)$ because 
$$\begin{array}{l}\sum\limits_{n=1}^{\infty} (ed|S|)^n\exp\left[\left(-\dfrac{\tau}{d+1}
+\dfrac{1}{10}\right)n\right] 
\le\dfrac{1}{10}\end{array}$$
whenever \ $(ed|S|)\exp\left(-\dfrac{\tau}{d+1}+\dfrac{1}{10}\right) 
< \dfrac{1}{11}$, \ i.e., \ $\tau > \tau_0$. Consequently, 
for $\pi=[\theta_i^{\alpha_i}]$ and  $\tau=\tau_0$ estimate (A.7) becomes $|r(\pi)| \le 
\exp\left(\log(13ed)\sum\limits_i \alpha_i |C_i| \right)$. Accordingly, for $\tau >\tau_0$ definition 
(A.5) implies an exponentially decaying bound $$|w(\pi)| \le \exp\left(-\dfrac{\tau-\tau_0}{d+1} 
\sum\limits_i \alpha_i |C_i| \right).$$

The Gibbs measure in a finite  volume $\Lambda$ with a boundary condition fixed on 
$\bbG\setminus \Lambda$ is defined by the probabilities of cylinder events (A.12) or 
equivalently (A.13) for the case of a single contour. For a general collection of contours 
$\{\theta_j\}$ expression (A.13) becomes 
$$\begin{array}{cl}
\mu_{\Lambda}(\{\theta_j\})&:=Z(\Lambda)^{-1}  Z(\Lambda \setminus F(\{\theta_j\})\; 
\prod\limits_j w(\theta_j)\\
\;& =\exp \left(- \sum\limits_{\pi \subseteq \Lambda:\  \pi \cap F(\{\theta_j\})\neq\emptyset} w(\pi)\right)\; \prod\limits_j w(\theta_j) , \end{array}\eqno (A.13.1)$$
where $F(\{\theta_j\})$ is the set of all contours $\theta'$ incompatible with at least one 
$\theta_j$. 

Consider a Van Hove sequence of  volumes $\Lambda_n \nearrow \bbG$, whose existence in 
$\bbG$ is ensured by the regularity requirement (c). Then the thermodynamic limit $n\to\infty$ 
of (A.13.1) is straightforward. This limit $\mu(\{\theta_j\})$ is given by the same expression (A.13.1) 
with a twofold difference affecting the sum in the exponent. 

First, all polymers containing a contour $\theta\subseteq \Lambda_n$ that is adjacent to 
$\bbG \setminus \Lambda_n$ are not included. The reason is that both the geometry and 
the statistical weight of such a contour are affected by the boundary condition, and therefore 
are $\Lambda_n$-specific. (Note that these $\Lambda_n$-specific statistical weights still satisfy 
(5) and, consequently, (A.3) implying the validity of (A.6).) 

Second, the condition $\pi \subseteq \Lambda_n$ is replaced with $\pi \subset \bbG$, i.e., 
the sum includes polymers which are not confined into $\Lambda_n$. The statistical weights 
$w(\theta)$ of contours constituting such polymers are calculated in a $\Lambda_{n'}$ 
containing $\theta$ and having ${\rm dist}\left(\theta,\bbG\setminus\Lambda_{n'} \right) \ge 2$. 
By construction, such a statistical weight does not depend on~$\Lambda_{n'}$. 

Thus, the discrepancy between the finite-volume and the infinite-volume versions of the sum 
in (A.13.1) is given by the polymers stretching from $\{\theta_j\}$ to $\bbG\setminus \Lambda_n$.
We already verified that the absolute value of the statistical weight of a polymer decays
exponentially with the size of the polymer support. Hence, the above discrepancy tends to $0$ 
as $n \to \infty$. The exponential decay of $\mu(\cdot)$-correlations follows from (A.13.1) in a 
similar way. The uniqueness of $\mu(\cdot)$ is a consequence of its independence of boundary 
conditions. 
\end{proof1}

\section{Partitioning a P3 tiling}\label{Section3}

We assume that the reader is familiar with the definition of a Penrose P3 tiling \cite{[P]} and the  
related terminology \cite{[GS], [Se]}. For the sake of a minimal readability we recall that P3 tilings are 
defined as the tilings of $\bbR^2$ by decorated thick and thin rhombi. The non-empty intersection of two 
rhombi is either a single vertex or a single edge. In the latter case their decorations must 
match. Each participating rhombus is obtained from the one of the prototypical two by shifts and 
rotations; reflections are not allowed as they change the orientation. Furthermore, we are 
mainly dealing with the so-called geometrical tilings \cite{[FL]} for which the decorations are removed, 
and we treat these tilings as planar graphs.

%The paper [FL] also can serve as a good source of the standard terminology with appropriate references to the original papers. 
The work \cite{[FSP]} analyzes a number of related phenomena, and similarities between the current paper and \cite{[FSP]} are not accidental. 
%It also contains an extensive bibliography.

\begin{figure}[H]
\begin{center}
\includegraphics[scale=0.9]{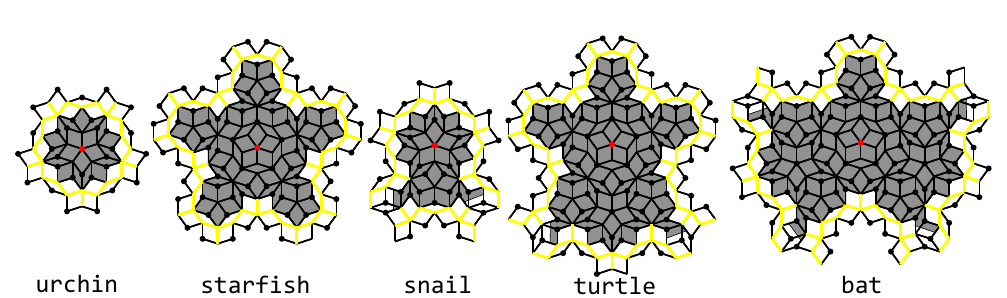}

\caption{The five basic patterns}
\end{center}
\end{figure}

Our construction is based on the five basic collared patterns shown in Figure 3; the proposed
names are given for presentational convenience and do not have biological meaning. A connected gray region in Figure 3 is referred to as the interior of a given pattern. The 
union of the interior and 
the  edge-adjacent white rombi forms a complete pattern. Accordingly, the edges of the white rhombi which are not shared with the interior are colored yellow to indicate the pattern boundary. Some yellow edges are colored twice, i.e., they are shared by two white rhombi 
that are edge-adjacent to 
the interior. The remaining white rhombi originate from neighboring patterns and constitute the collar which ensures  recognizability of a given pattern. If two patterns are adjacent in a tiling, then their interiors do not intersect. Correspondingly, the sets of vertices belonging
to the closures of these interiors also do not intersect. The (occupied) red vertices are referenced as the centers of the corresponding patterns. The vertices of the collar which are at the graph distance $1$ from the interior constitute the statistical-mechanical boundary of the pattern. Only particles  located at this boundary can affect the configuration of particles inside the pattern interior. The black dots indicate the perfect configuration inside the pattern interior. The black dots at the boundary represent one example of several possible statistical-mechanical boundary configurations.

%\vfil\eject

%\centerline{%
% \pdfximage height 140px {Patterns.pdf}%
%  \pdfrefximage\pdflastximage}
%{\ninerm \hskip5px urchin \hskip50px starfish \hskip60px snail \hskip60px turtle \hskip95px bat}

%\vskip .5 cm  
%\centerline{\ninerm Figure 3. The five basic patterns}

%\vskip .5 cm  \noindent
%\centerline{%
% \vbox{%
%    \hsize=0.8\hsize   % 60% of page width (adjust as needed)
%\noindent
%\vfil\eject

\begin{lem-1}\label{lem:1}
For each of the patterns, independently of an admissible configuration of particles at its statistical-mechanical boundary, the perfect configuration shown in {\rm Figure~3} contains at least one more particle than any other admissible configuration in the pattern interior.
\end{lem-1}

\begin{proof1}
For each pattern all vertices of the corresponding graph, except for the central vertex, can be decomposed into the union  of several loops of even length such that in each loop the perfect configuration occupies exactly a half of the vertices.

\begin{figure}[htbp]
\begin{center}
\includegraphics[scale=0.9]{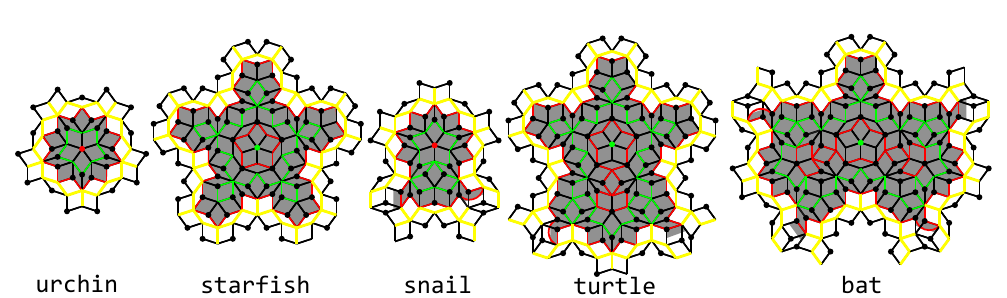}

\caption{The five basic patterns with loops}
\end{center}
\end{figure}

%\centerline{%
% \pdfximage height 140px {PatternsLoops.pdf}%
% \pdfrefximage\pdflastximage}
%{\ninerm \hskip5px urchin \hskip50px starfish \hskip60px snail \hskip60px turtle \hskip95px bat}

%\vskip .5 cm  
%\centerline{\ninerm Figure 4. The five basic patterns with loops}

\noindent
In Figure~4 the alternating (possibly, self-intersecting) loops are shown in green and red, respectively. It is clear that at 
most a half of a loop can be occupied. Hence, no configuration inside a pattern other than 
the perfect one can have more particles. The particle at the center  is the one that gives an
advantage to the perfect configuration. 
\end{proof1}

%Lemma 2 below establishes that the entire set of vertices in a P3 tiling is essentially partitioned
%into the interiors of patterns. As a consequence of the lemma, the union of configurations 
%maximizing the number of particles in each pattern provides the global configuration of {\color{blue}the ???} 
%maximal density. The precise meaning of the term essentially is that two different patterns may 
%have two vertices in common (which creates a dependence between them). However, the densest 
%configuration inside each of {\color{blue}the ???} two patterns does not contain particles at these vertices. Thus, the 
%existing local dependence does not affect the globally densest configuration. The locally densest 
%configurations emerging from Lemma~1 turn out to be not contradicting each other.

\begin{lem-2}\label{lem:2} 
The vertex set of a  P3 tiling  %$\bbG$ %of the entire $\bbR^2$
is uniquely partitioned into the sets of vertices belonging to the interiors of patches corresponding to the five basic patterns. The emerging structure inherits the self-similarity of the underlying P3 tiling. In particular, the centers of the patches and the links that connect the centers of adjacent patches form a supertiling that is MLD with the underlying~P3 tiling.
\end{lem-2}

\begin{proof1} First, we partition P3 into level-$4$ supertiles (cf. \cite{[AP]}, \cite{[LMS]}) which are images of 
four times substituted original thick and thin rhombi.

\begin{figure}[H]
\begin{center}
\includegraphics[scale=0.44]{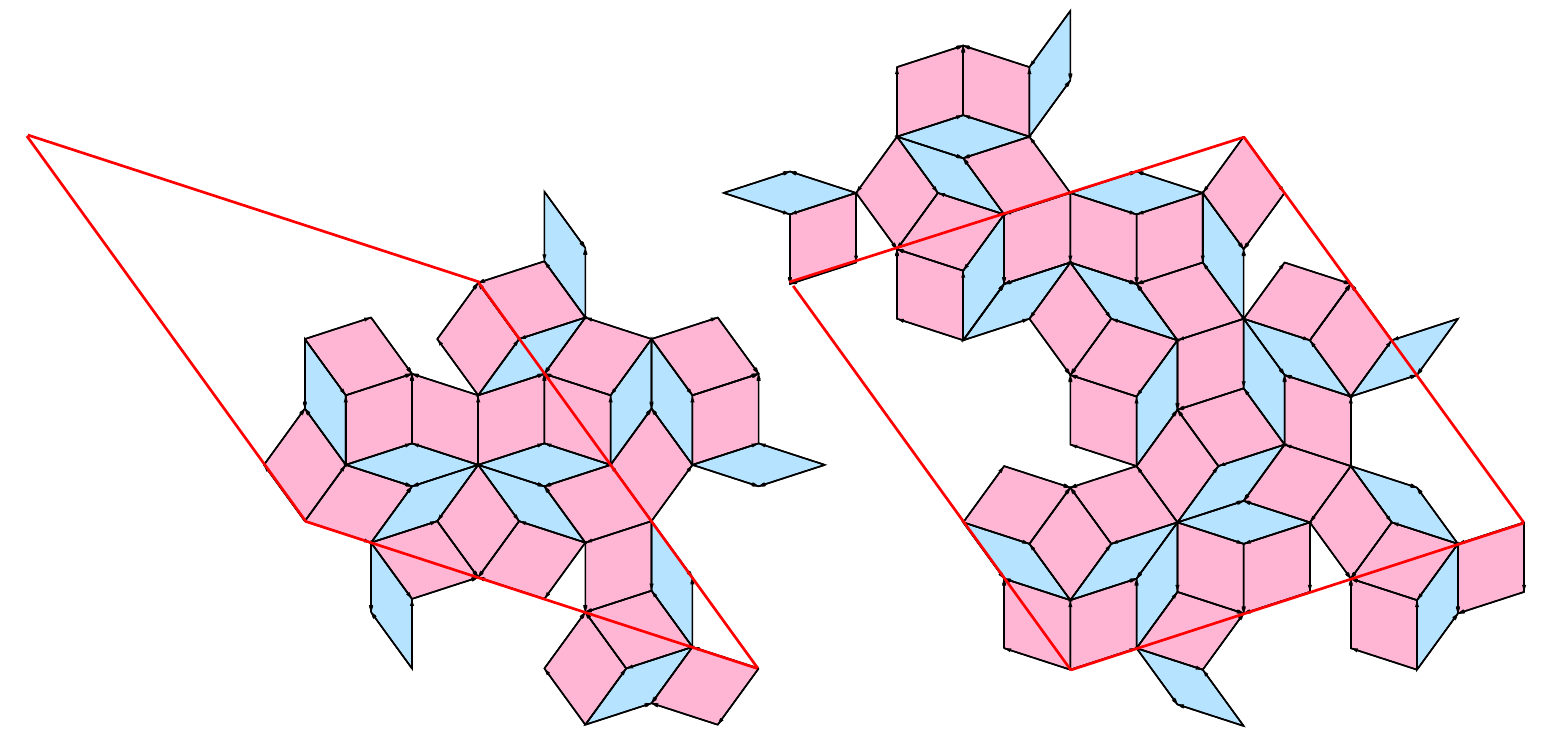}
\caption{The level-$4$ supertiles}
\end{center}
\end{figure} 

Figure 5 shows a fully decorated result of the fourth iteration of the standard substitution dynamics applied to the thin and thick unit rhombi, respectively. In the course of a single substitution step, a thin rhombus generates one thin and one thick rhombus while a thick 
rhombus generates one thin rhombus and two thick ones.

The boundaries of adjacent supertiles are matching each other and therefore serve as natural decorations (replacing the original single and double arrows). The boundaries in Figure~5 deviate too far from the scaled rhombi they  represent (outlined in red in Figure~5). It is more convenient to use an equivalent pair of supertiles which is presented in Figure 6. One can see that these supertiles also have boundaries which serve as rhombus decorations and must be matched exactly between the adjacent rhombi.

\begin{figure}[H]
\begin{center}
\includegraphics[scale=0.86]{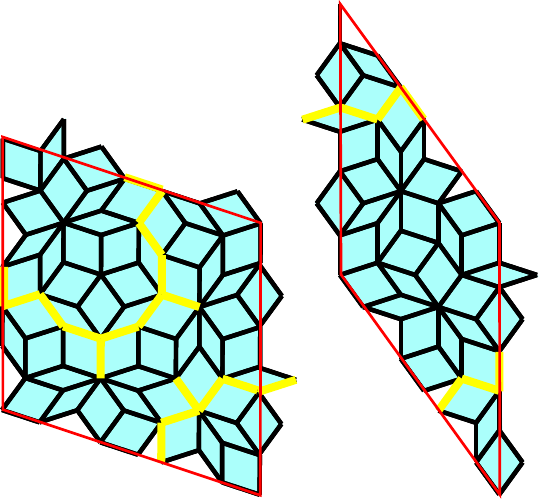}

\caption{The level-$4$ equivalent supertiles}
\medskip
The yellow edges form the boundaries of the patches.
 \end{center}
\end{figure}
\noindent
The supertiles in Figure 6 form a $\varphi^4$ times scaled version of a P3 tiling. An immediate consequence is that the yellow edges form non-branching broken lines. 

The extended 
$0$-atlas of a P3 tiling is presented in the Figure~7, together with the
frequencies of its elements (see Figures~6~and~8 in \cite{[FL]} and FIG.~3 in \cite{[FSP]}).

\begin{figure}[H]
\begin{center}
\includegraphics[scale=0.85]{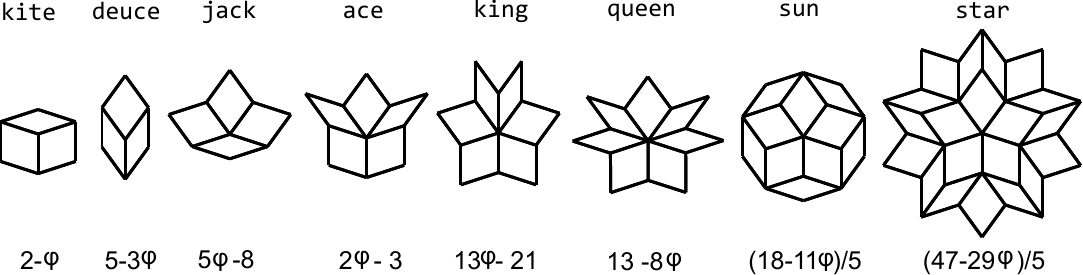}

\caption{The extended $0$-atlas of a geometrical P3 tiling}
\end{center}
\end{figure} 

Applying the extended  $0$-atlas to a P3 supertiling constructed from supertiles in Figure~6, it is not hard to conclude that the supertile vertices are one-to-one mapped to the $1$-corona and $2$-corona in the original P3 tiling, which are shown in Figure 8. 

\begin{figure}[h]\label{Fig.8}
\begin{center}
\includegraphics[scale=1]{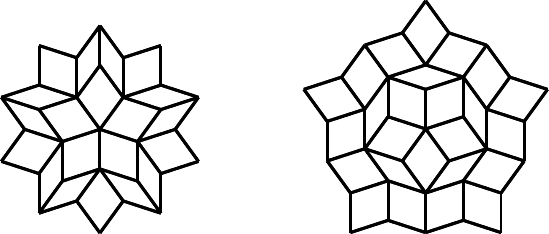}

\caption{The coronas in a P3 tiling}
 \end{center}
\end{figure}

Furthermore, the kite element from the supertiling extended $0$-atlas is the only one whose center does not correspond to a patch center. The center of a deuce element from the extended $0$-atlas is always the center of a ``bat''. The center of a jack element from the extended $0$-atlas is always the center of a ``turtle''. The center of an ace element from the extended $0$-atlas is always the center of a ``snail''. The center of a king, queen or star from the extended $0$-atlas is always the center of an ``urchin''. Finally, the center of a sun element from the extended $0$-atlas is always the center of a ``starfish''. 

Note that a patch with a given center sometimes goes outside of the supertiling $0$-atlas but is always confined to the supertiling $1$-atlas (cf. Figure~8 in \cite{[FL]}).

Based on the emerged scaled P3 supertiling, we define a different supertiling consisting of rhombi, kites (different from ones in Figure~7), and two reflection symmetric trapezes which we call an RKTT supertiling. The RKTT supertiling is obtained from the P3 supertiling by removing all vertices which are the centers of all $0$-atlas elements of the kite type. More specifically, the edges which are incident to a removed vertex are shared either  by thin and thick rhombi or by two thick ones. The two kite edges which are shared with thin rhombi are completely deleted from the graph, while the kite edge shared by two thick rhombi is prolonged to join the nearest P3 vertex. Figure~9 illustrates the result of this operation. 

\begin{figure}[H]\label{Fig.9}
\begin{center}
\includegraphics[scale=0.65]{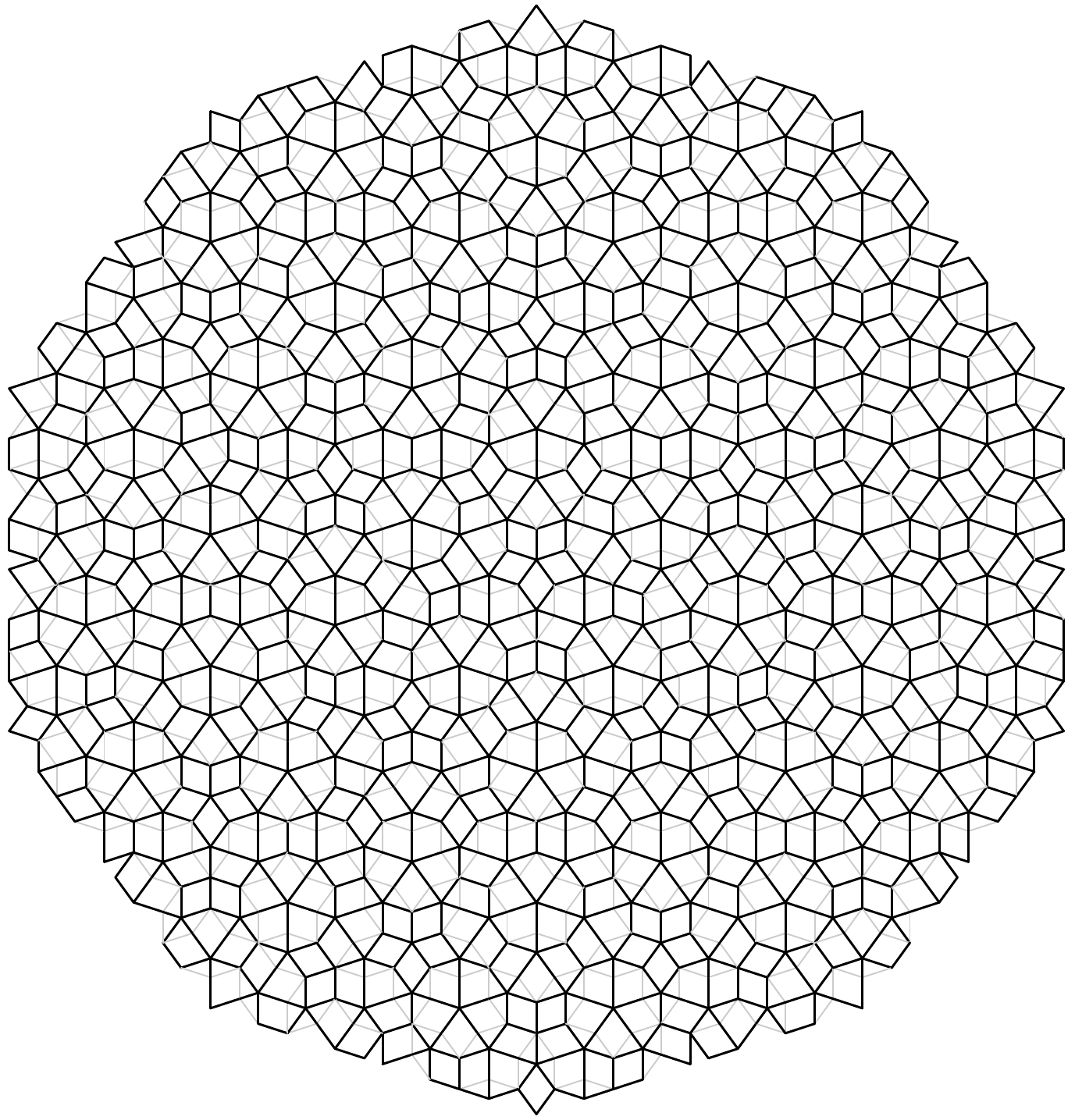}
\end{center} 
\caption{Superimposed P3 and RKTT tilings}
\bigskip
The RKTT tiling obtained from a P3 tiling by adding all diagonals of the thin rhombi and joining them with the adjacent collinear unit edges into new, longer edges. Simultaneously, the thin rhombi edges which are incident to the corresponding joining vertex are not included into the RKTT and consequently are painted light gray.

\end{figure}

\begin{figure}[H]\label{Fig.10}
\begin{center}
\includegraphics[scale=0.65]{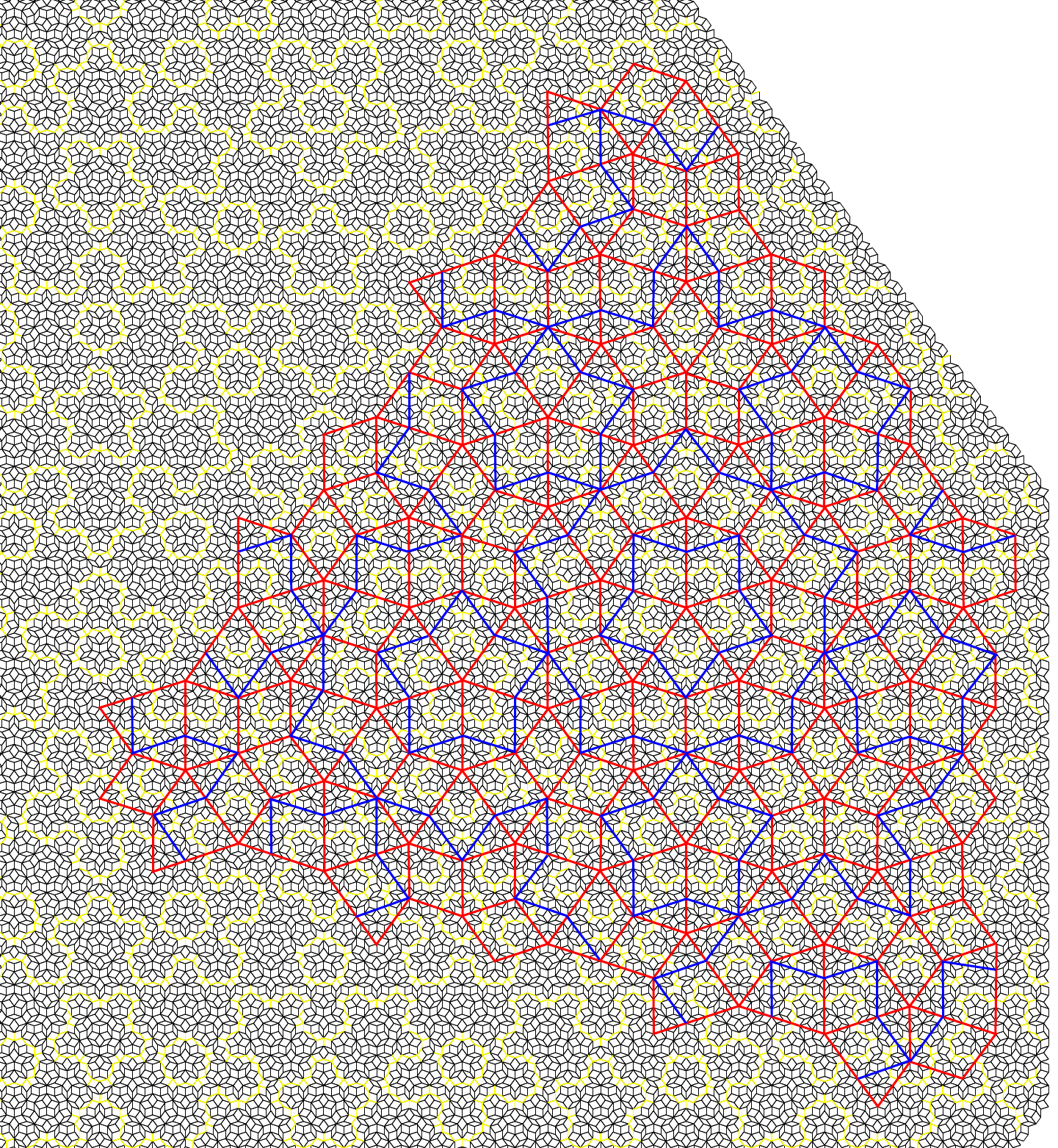}

\caption{ P3 and RKTT supertilings and patches with boundaries}
\end{center}
\end{figure}

Figure~10 presents a large patch of a 5-fold symmetric P3 tiling, together with both  supertilings (P3 and RKTT) and yellow separating boundaries. These yellow lines partially separate different patches leaving some ``snails'', 
``turtles'' and ``bats'' not separated yet. The cuts separating such ``snails'', 
``turtles'' and ``bats'' should be done along the narrow bridges where two yellow boundaries 
are getting closest to each  other (see the central thick rhombus in Figure~6).

The supertile 
vertices and corresponding straight edges (see Figure~6) are shown inside a $\pi /5$ 
angle only and can be extended via a 5-fold rotational symmetry. The edges of the P3 supertiling are colored blue, the edges of the corresponding RKTT supertiling are colored red, with some blue edges getting hidden under the red ones.
%The blue edges are from the scaled P3 but not from the scaled RKTT. Some of the red edges belong to both the scaled P3 and the scaled RKTT. The scaled RKTT is given by the red edges only. 
Each red RKTT star is unambiguously mapped into a basic pattern. Under this mapping the center of the star becomes the center a basic pattern (cf. Figure~3). This mapping can be described as follows. 

Any RKTT star consisting of five kites corresponds to a ``starfish''. Any RKTT star consisting of five rhombi corresponds to an ``urchin''. Other RKTT stars corresponding to an ``urchin'' are either the ones consisting of two rhombi and two trapezes or the ones consisting of one kite and two trapezes. Any RKTT star consisting of two rhombi and a kite corresponds to a ``snail''. Any RKTT star consisting of one kite and four trapezes corresponds to a ``turtle''. Any RKTT star consisting of one rhombus, two kites and two trapezes corresponds to a ``bat''. 

These observations lead to the unique partition of a P3 tiling which is constructed in the following way:

(I) the original P3 tiling is uniquely partitioned into supertiles, 

(II) these supertiles are uniquely regrouped into RKTT supertiles,

(III) the corresponding RKTT stars are uniquely mapped into five basic patters,

(IV) the union of the resulting patches centered at the RKTT vertices cover the entire original P3 tiling.

\noindent
The last fact is again verified directly and implies that there is no rhombus in the original P3 tiling which does not belong to one of the five basic patterns. 

Each step of the above construction produces either a tiling or a supertiling that is MLD with the previous one. Therefore, the entire construction inherits the self-similarity of the original P3 tiling.
\end{proof1}

\begin{rc} {\rm For the 5-fold symmetric P3 tiling which is the fixed point of the substitution dynamics, one can specify a two-symbol alphabet and a word-substitution rule describing how the next closed yellow loop encircling the origin is derived from the previous one. The presence of an infinite family of concentric yellow loops implies that any yellow broken line in this P3 tiling has a finite length and therefore is a loop. Consequently, any connected component of ``snails'', ``turtles'' and ``bats'' is finite. A generic P3 tiling can have only a finite number of infinite yellow broken lines (upper-bounded by an absolute constant). However, for any P3 tiling the thickness of the layer between any two adjacent yellow broken lines (finite or infinite) is not larger than the diameter of the ``bat'' pattern.}
\end{rc}

\section{Proof of Theorem}\label{Proof}

Lemmas 1 and 2 provide the structure of the unique ground state (perfect configuration) of the model (see Figure 2). The corresponding graph-density or, equivalently, the maximal graph-density of an independent set in a P3 tiling can be calculated via the frequencies of the $0$-atlas and $1$-atlas elements in a geometrical P3 tiling established in \cite{[FL]} (cf. Figure~6~and~8). From Figure 3 we conclude that the maximal number of particles inside a pattern is $16,51,23,63,75$ for the ``urchin'', ``starfish'', ``snail'', ``turtle'' and ``bat'', respectively. Consequently, the  ground state particle density is equal to
$$\begin{array}{l}
\dfrac{1}{\varphi^8}\bigg[75 (5-3\varphi) + 63(5\varphi-8) + 23(2\varphi-3)
 +16(13\varphi-21)\\ \qquad + 16 (13-8\varphi) 
+ 16\left(\dfrac{47}{5}-\dfrac{29}{5}\varphi\right)+ 51\left(\dfrac{18}{5}
-\dfrac{11}{5}\varphi\right)\bigg]\\
\hskip 5.3cm =\dfrac{1}{2} (57 - 25 \sqrt{5}) \approx 0.54915,\end{array} \eqno{(6)}$$
where the factor ${\varphi}^{-8}$ accounts for the fact that we are working in the level-4 supertiling. This establishes the first statement in the theorem. 

From the Gibbs measure perspective, Lemmas 1and 2 provide the reduction of the original model with Hamiltonian 
(1) to the simplified model (3) discussed in Proposition. More precisely, the RKTT graph is the 
desired graph $\bbG^\rST $ with  the vertex degree at most five. For any $\bv \in \bbG^\rST $ 
the spin space $S_{\bv}$ is the set of admissible configurations in the corresponding
pattern, denoted as $P_{\bv}$. Given adjacent vertices $\bu,\bv \in \bbG$,
the space $S_{\bu\bv}$ is the set of admissible configurations in the union $P_{\bu} \cup P_{\bv}$.
(Note that $S_{\bu\bv}$ is not $S_\bu\times S_\bv$ owing to the hard-core exclusion,
in agreement with the assumptions of Proposition.) Accordingly, 
the value $U\big(\sigma(\bv)\big)$ is the negative logarithm of the product of activities of the particles in the corresponding configuration $\phi(P_{\bv})$. A particle shared by two patterns contributes $\sqrt{ u}$ into the statistical weight of each pattern. The remaining conditions of Proposition are verified in a straightforward manner. Namely, $\tau=\log( u)$, $d=5$, $|S| \le 2^{150}$ implying 
that ${\ov u} = e^{10 \cdot 6}\cdot 2^{150}\cdot e\cdot 5$. A better (though still not optimal) estimate can be obtained in the following way.

For a simple loop of length $2m$ in $\mathbb{G}$ the corresponding partition function is equal to 
$$2^{-2m} \left((1+\sqrt{1+ 4 u})^{2m} +  (1-\sqrt{1+4 u})^{2m} \right) < 2(1+\sqrt{ u})^{2m}.\eqno{(7)}$$
Our patterns are unions of at most three loops; cf. Lemma~1 and Figure 4. Therefore, the partition function 
over all admissible configurations, but the ground state, inside a pattern containing $2n+1$ 
vertices does not exceed $2\cdot 2^3\cdot (1+\sqrt{ u})^{2n}$ where the factor $2$ accounts
for the possibility to have the pattern center occupied or vacant. Thus, the sum of statistical weights of all contours in $\bbG^\rST$ whose support covers the origin does not exceed
$$\sum_{k=1}^{\infty} (5^3e)^k\left(16(1+\sqrt{ u})^{148} \over  u^{75} \right)^k <  \sum_{k=1}^{\infty} (2000e)^{k}\left(e^{148/\sqrt{ u}}\over  u\right)^k. \eqno{(8)}$$
Here the factor $(5^3e)^{k}$ is the standard upper estimate for the number of $3$-connected subsets in graph $\bbG^\rST$ (of maximal degree $5$) originating at a given graph vertex. These subsets are assumed to represent all vertices with $\phi(\bv) \not = s_{\bv}$ inside a contour. The second factor in the sum on the LHS of~(8) is the upper bound of the statistical weight of such a contour. The quantity in the RHS is sufficiently small for $u>100000$ to ensure the validity of condition (A.3) in \cite{[KLMS]} and, consequently, the convergence of the polymer expansion~(A.4). This completes the proof of the theorem.

\newpage
\noindent
{\bf Acknowledgments}

\vskip .5 cm  
The authors are thankful to ChatGPT for the crash course in Penrose tilings, pointing towards relevant references, and Java coding.

The authors thank Tyler Helmuth and Senya Shlosman for a number of critical remarks and
suggestions for improving an initial version of the manuscript.

IS and YS thank Math Department, PSU, for support. 

YS is grateful to DPMMS, University of Cambridge, and St John's College, Cambridge, for the continuing support. YS thanks Prof. Phan Th\`anh Nam
and Mathematics Institute, LMU Munich, for hospitality and support during a visit
in June, 2024.

\vfil\eject
%\bigskip
\noindent
{\font\big=cmb10 at 13pt \big Addendum}

\vskip .5 cm  
We attach to the paper a collection of stand-alone Java utilities used to generate large scale 
SVG images included into the text.

The {\tt PenroseP3SunSVG.java} is used to draw an undecorated 5-fold symmetric P3 tiling generated via $k$ times iterated Robinson triangle substitution, where $k$ is an input parameter (see Figure~1).

The {\tt PenroseP3SunSVGRKTT.java} is used to draw an RKTT tiling atop of an originating undecorated 5-fold symmetric P3 tiling generated via $k$ times iterated Robinson triangle substitution, where $k$ is an input parameter (see Figure~9).

The {\tt PenroseP3\_ABC\_Coloring.java} is used to draw the ground state configuration in a 
5-fold symmetric P3 tiling generated via $k$ times iterated Robinson triangle substitution, 
where $k$ is an input parameter. Particles of different parities are shown in red and blue colors. The edges without particles are shown in yellow. The graphics parameters like colors, thickness, diameters are easily adjustable via self-explanatory named variables (see Figure~2 and 
Figure~10).

The {\tt PenroseP3.java} is used to draw a P3 tiling generated from an arbitrary seed patch is 
$k$ times iterated original Penrose rhombi substitution, where $k$ is an input parameter. An
arbitrary initial seed patch can be specified inside the {\tt seedOneThick} or {\tt seedOneThinFromUserExample} methods (see Figure~5).

%\end{plain}
\end{document}